\documentclass[aps,showpacs, nofootinbib,floatfix]{revtex4}
\usepackage{graphicx, graphics, color}

\input amssym.tex
\input amssym.def

\newcommand{\be}{\begin{equation}}
\newcommand{\ee}{\end{equation}}
\newcommand{\ben}{\begin{eqnarray}}
\newcommand{\een}{\end{eqnarray}}

\newcommand{\la}{{\lambda}}

\newcommand{\si}{{\sigma}}

\newcommand{\na}{\nabla}


\begin{document}

\title{Can Dirac fermions destroy Yang-Mills black hole?}
%%%%%%%%%%%%%%%%%%%%%%%%%%%%%%%%%%%%%%%%%%%%%%%%%%%%%%%%%%%%%%
%\author{Gary W.Gibbons}
%\affiliation{DAMTP, Centre for Mathematical Sciences,  \protect \\
%University of Cambridge\protect \\
%Wilberforce Road, Cambridge, CB3 0WA, UK\protect \\
%g.w.gibbons@damtp.cam.ac.uk} 

\author{{\L}ukasz Nakonieczny and Marek Rogatko}
\email{rogat@kft.umcs.lublin.pl, 
marek.rogatko@poczta.umcs.lublin.pl,
lnakonieczny@kft.umcs.lublin.pl}
\affiliation{Institute of Physics \protect \\
Maria Curie-Sklodowska University \protect \\
20-031 Lublin, pl.~Marii Curie-Sklodowskiej 1, Poland }

%\author{Marek Rogatko}
%\affiliation{Institute of Physics \protect \\
%Maria Curie-Sklodowska University \protect \\
%20-031 Lublin, pl.~Marii Curie-Sklodowskiej 1, Poland \protect \\
%rogat@kft.umcs.lublin.pl \protect \\
%marek.rogatko@poczta.umcs.lublin.pl}

%%%%%%%%%%%%%%%%%%%%%%%%%%%%%%%%%%%%%%%%%%%%%%%%%%%%%%%%%%%%%%%%%%%%
\date{\today}
%\pacs{04.30.Nk, 04.40.-b}

%%%%%%%%%%%%%%%%%%%%%%%%%%%%%%%%%%%%%%%%%%%%%%%%%%%%%%%%%%%%%%%%%%%%%%%%%%%%%%%%%%%%%%%%%%%%%%%%%%%
\begin{abstract}
We study the four-dimensional Einstein-Yang-Mills black hole in the presence of Dirac fermion field.
Assuming a spherically symmetric static asymptotically flat black hole spacetime we consider
both massless and massive fermion fields. The $(4 + 1)$-dimensional 
Einstein-Yang-Mills system effectively reducing to Einstein-Yang-Mills-Higgs-dilaton model,
was also taken into account.
One finds that fermion vacuum leads to the destruction of black holes in question.

\end{abstract}
%%%%%%%%%%%%%%%%%%%%%%%%%%%%%%%%%%%%%%%%%%%%%%%%%%%%%%%%%%%%%%%%%%%%%%%%%%%%%%%%%%%%%%%%%%%%%%%%%%%%

\maketitle

%%%%%%%%%%%%%%%%%%%%%%%%%%%%%%%%%%%%%%%%%%%%%%%%%%%%%%%%%%%%%%%%%%%%%%%%%%%%%%%%%%%%%%%%%%%%%%%%%%%%%
\section{Introduction and notation.}
Gravitational collapse is one of the most important issue of general relativity and its extension to higher 
dimensional spacetime connected with M/string theory schemes of unification of the all known forces of Nature.
One expects that a newly born black hole emerging from the gravitational collapse of a massive star will settle
down to the stationary axisymmetric or static spacetime. The uniqueness theorem (or classification) of nonsingular black 
hole solutions states that a stationary axisymmetric solution of Einstein-Maxwell (EM) electrovacuum 
equations is isometrically diffeomorphic to the domain of outer communication of Kerr-Newman spacetime 
\cite{uniq}. 
\par
On the other hand, the complete classification of $n$-dimensional charged black holes both with non-degenerate and 
degenerate component of the event horizon was given in Refs.\cite{nd}.
Partial results for the very nontrivial case of $n$-dimensional rotating black hole uniqueness theorem were 
provided in \cite{nrot}. The aforementioned studies comprises also the case of extremal axisymmetric black hole both in 
EM theory and the low-energy limit of the string theory (the so-called EMAD-gravity) and supergravities theories
\cite{sugra}.\\
As far as the uniqueness theorem of non-Abelian black holes is concerned, the situation is far more complicated
(see \cite{heu98} and references therein). It turned out that any static solution of Einstein-Yang-Mills (EYM)
equations ought to either coincide with Schwarzschild one or posses some non-vanishing Yang-Mills (YM) charges. But it 
happened not to be the case when static black hole solutions with vanishing charges were discovered \cite{staticym}.
They were asymptotically indistinguishable from Schwarzschild black hole. Moreover, in Ref.\cite{rid95}
it was shown that static black hole of the considered class with {\it magnetic} charge needed not even be axially symmetric.
In the light of the above, one can remark that the non-Abelian black holes reveal considerably more composed structure
comparing to the EM ones.
\par
Recently, studies of fermions in various backgrounds attract more attention. Exact solutions of the 
Dirac equation in curved spacetime may be a useful tool for investigations of physical properties
of the considered spacetimes. Dirac fields were elaborated \cite{gib93}
in the context of EYM
background found in Ref.\cite{bar88}, in the near-horizon limit of Kerr black hole \cite{sak04},
in Bertotti-Robinson spacetime \cite{br}, in spacetimes of black holes with nontrivial topology of the event horizon
\cite{goz10}, in the vicinity of black holes with topological defects \cite{nak11} and in the spacetime
of black string \cite{nak12}. The intermediate and late-time decays of 
massive Dirac fermions in various black hole spacetimes were also elaborated
\cite{jin04}-\cite{br08}.
\par
The other tantalizing problem is the behaviour of black holes 
and the surrounding matter fields. Depending on the matter model in question, black holes 
may allow to exist the nontrivial fields outside the event horizon.
Are there any configurations of matter fields that can destroy the
emerging black hole? 
This question was tackled in Refs.\cite{haj74}.
In Refs.\cite{loh84}
it was revealed that Reissner-Nordstr\"om (RN) solution with both an electric and magnetic charges can be destroyed in the presence
of a massless Dirac fermion field. On the other hand, it was revealed \cite{smo} that the only
black hole solutions of four-dimensional spinor Einstein-dilaton-Yang-Mills field equations
of motion were those for which spinors vanished identically outside black hole. It means that Dirac fermion
fields either enter the black hole in question or escape to infinity.
Recently, it was shown \cite{shi06}
that matter configuration composed of
a perfect fluid could not be at rest outside a four-dimensional black hole in asymptotically flat 
static spacetime.
\par
In our paper we use the bosonozation technique by which 
the fermionic degrees of freedom can be described by a scalar field.
We shall elaborate the problem of the influence of Dirac fermion fields on Yang-Mills (YM) black holes.
In our considerations
we assume that the black hole in question is spherically symmetric and static. We take into account
the ordinary four-dimensional static spherically symmetric asymptotically flat EYM black hole and the
four-dimensional Einstein-Yang-Mills-Higgs-dilaton (EYMHd) system deduced from five-dimensional EYM model
while the Dirac fermion field will be treated in a s-wave sector.\\
The organization of this paper is as follows.
In Sec.II we briefly review the basic facts concerning four-dimensional EYM black hole and Dirac fermion
system. Then, we consider the massless as well as massive fermions case and their influence
on black hole. Their asymptotical behaviours will be also discussed. Sec.III will be devoted to
the backreaction process of Dirac fermions on YM fields. In Sec.IV we shall elaborate
$(4+1)$-EYM system which effectively reduces to EYMd field equations, where
both the matter fields and the line element coefficients do not depend on the fifth dimension.
We conclude our investigations in Sec.V.

%%%%%%%%%%%%%%%%%%%%%%%%%%%%%%%%%%%%%%%%%%%%%%%%%%%%%%%%%%%%%%%%%%%%%%%%%%%%%%%%%%%%%%%%%%%%%%%%%%%
%%%%%%%%%%%%%%%%%%%%%%%%%%%%%%%%%%%%%%%%%%%%%%%%%%%%%%%%%%%%%%%%%%%%%%%%%%%%%%%%%%%%%%%%%%%%%%%%%%%
\section{Four-dimensional Einstein-Yang-Mills Black Hole and Dirac fermions}
In this section we shall pay attention to static spherically symmetric solution
of EYM field equations. In the case under consideration the line element can be provided by
\be
ds^2 = -A^2(r)~dt^2 + B^2(r)~dr^2 + C^2(r)~d\Omega^2, 
\ee
where $d\Omega^2$ is a metric on $S^2$-sphere. In what follows it will be convenient to
introduce the {\it tortoise coordinate} defined as $ dr_{*} = \frac{ B }{ A } dr$.
Just the underlying metric yields
\be
ds^2 = -A^2(r)~dt^2 + A^2(r)~dr_{*}^2 + C^2(r)~d\Omega^2. 
\ee
The main topic of our research will be the influence of Dirac fermion fields on EYM black hole.
Fermions under consideration will be described by the Dirac equation provided by
\be
i\gamma^{\mu}D_{\mu} \psi - m\psi = 0,
\ee
where the covariant derivative $D_{\mu}$ implies
\be
D_{\mu} = \nabla_{\mu} - i \lambda H_{\mu}.
\ee
$\la$ is the gauge coupling constant, while the component of Yang-Mills field have the forms as
\be
H_{\mu} = e_{\mu}^{i} H_{i}, \qquad H_{i} = a_{i} {n}^k ~{\tau}_k + 
\frac { 1 - K(r) }{ 2 \lambda C } \epsilon_{i j k}~ n^{j}~ \tau^{k},
\ee 
where  $a_{i}$ is the {\it electric} and $K$ the {\it magnetic} part of Yang-Mills vector.
$n_a$ is the unit normal vector, while $\tau_a$ is a generator of $SU(2)$ group. On the other hand,
$e^{i}_{\mu}$ are basis one-form defined by $g_{\mu \nu} = e^{i}_{\mu} e^{j}_{\nu} \eta_{i j}$, where
$\eta_{i j}$ is metric tensor for Minkowski spacetime.  
The gamma Dirac matrices in a flat spacetime are defined by the relations
\be
\gamma^0 = \pmatrix{ 0&1 \cr 1&0}, \qquad \gamma^i = \pmatrix{ 0& \sigma^i \cr - \sigma^i &0}.   
\ee
where $\sigma^i$ are the usual Pauli matrices. It turned out that the Dirac operator takes the form as
\be
\gamma^{\mu}D_{\mu} = \pmatrix{ 0& D^{+} \cr D^{-}&0}, 
\ee
where by $D^{\pm}$ we have denoted the following relation:
\ben
D^{\pm} &=& A^{-1}\partial_{t} - 
i \lambda \{ [ \sigma^{0}a_{0} \pm \sigma^{1}a_{1} ] \bar{n} \cdot \bar{\tau} \pm 
\frac{ K - 1}{ 2 \lambda C } \bar{n} \cdot \bar{\sigma} \times \bar{\tau} \} + \nonumber \\
&\pm&  \bar{\sigma} \cdot \bar{n} A^{-1}\partial_{r_{*}} 
\pm  \bar{\sigma} \cdot \bar{n} \{ A^{-1} C^{-1} \partial_{r_{*}} C +  
\frac{ 1 }{ 2 } A^{-1} A^{-1} \partial_{r_{*} }A  \}
\pm  C^{-1} D_{S^2}. 
\een
Because of the fact that we restrict our attention to the $s$-wave sector, it can be spanned by
two states $\chi_{1}$ and $\chi_2 = \si^a~n_a~\chi_1$ being the {\it hedgehog} spinor ansatz
\cite{hedhog}.
Moreover, they will obey the properties as
\ben
\bar{\sigma} \cdot \bar{n} \chi_{1/2} &=& \chi_{2/1}, \qquad D_{S^2} \chi_{1/2} = \mp \chi_{2/1}, \\
(\bar{n} \cdot \bar{\sigma} \times \bar{\tau}) \chi_{1/2}  &=& \mp 2 i \chi_{2/1}, 
\qquad  \bar{n}( \bar{\sigma} + \bar{\tau}) \chi = 0.
\een
In terms of its components, 
spinor $\psi$ can be written as
\be
\psi = \pmatrix{ \psi_{L} \cr \psi_{R} }, \qquad \psi_{L/R} = f_{L/R}(r_{r_{*}} ,t) \chi_{1} + g_{L/R} \chi_{2}.
\ee
By virtue of the above, the
Lagrangian for two-component left and right-handed spinors implies
\be
\mathcal{L}_{F} = i\bar{\psi}_{R} D^{+} \psi_{R} + 
i\bar{\Psi}_{L} D^{-} \psi_{L} - m \bar{\psi}_{R} \psi_{L} - m \bar{\psi}_{L} \psi_{R}.
\ee
On this account we can integrate over the angular degrees of freedom in Dirac fermion action.
From now on we will work with curved 2-dim spacetime: $ds^2 = - A^2 dt^2 + A^2 d r_{*}^2$. Latin letters from 
the beginning  of alphabet will refer to the curved spacetime, i. e., $a = t, r_{*}$, while those from the end of 
the alphabet are bounded with 
the flat spacetime, i. e., $i = 0, 1$.
In this spacetime the nonzero components of the spin connection is $\omega^{0 1} = \partial_{r_{*}} \ln{A}~dt$, while
the covariant derivative of spinor field has the form $\nabla_{a} = \partial_{a} + \frac{ 1 }{ 2 } 
\omega^{ i j}_{ a } \sigma^{i} \sigma^{j}$. In our considerations, we introduce
two-dimensional spinors $F_{L/R} = \pmatrix{ f_{L/R} \cr g_{L/R}} $
and we use the symmetric form of the Dirac operator 
$ \gamma^{\mu} \stackrel{\leftrightarrow}{D}_{\mu} = \frac{1}{2} 
[  \gamma^{\mu} \overrightarrow{D}_{\mu} -  \gamma^{\mu} \overleftarrow{D}_{\mu} ] $. It enables us to 
rewrite action for fermion fields in the form
\be
S_{F} = 4 \pi \int dt \int A^2 \mathcal{L}^{2}_{F} dr_{*}, 
\ee
as well as the Lagrangian for them, which implies the following:
\ben
\label{psi-2dim}
\mathcal{L}^{(2)}_{F} = 
i~ C^2~ \bar{F}_{R} \stackrel{\leftrightarrow}{ D^{+} }_{(t,r_{*})} F_{R} - 
C^2~ \frac{ K }{ C }~ \bar{F}_{R}~ \sigma^{2}~ F_{R}
 &-&  C^2~ \lambda~ \bar{F}_{R}~ (\sigma^{i} a_{i})^{+}~ \sigma^{1}~ F_{R} + i~ C^2~ A^{-1}~\partial_{r_{*}} 
\ln(\sqrt{A} C )~ \bar{F}_{R}~ \sigma^{1}~ F_{R} \nonumber \\
+
i~ C^2~ \bar{F}_{L}~ \stackrel{\leftrightarrow}{ D^{-} }_{(t,r_{*})}~  
F_{L} + C^2~ \frac{ K  }{ C }~ \bar{F}_{L}~ \sigma^{2}~ F_{L}
&-&  C^2~ \lambda~ \bar{F}_{L}~ (\sigma^{i} a_{i})^{-}~ \sigma^{1}~ F_{L} - 
i~ C^2~ A^{-1}~\partial_{r_{*}} \ln(\sqrt{A} C )~ \bar{F}_{L}~ \sigma^{1}~ F_{L} \nonumber \\
&-& m~ C^2~ \bar{F}_{R}~ F_{L} - m~ C^2~ \bar{F}_{L}~ F_{R},
\een
where we have denoted by
$D^{ \pm }_{( t, r_{*})} = A^{-1} [ \sigma^{0} \partial_{t} \pm \sigma^{1} \partial_{r_{*}}]$ and 
$(\sigma^{i}a_{i})^{\pm} = \sigma^{0}a_{0} \pm \sigma^{1}a_{1}$.
We are now in a position to approach the question of the equations of motion for the above system.
Namely, the Dirac equations for the two-dimensional fermions are provided by the relations
\ben
\label{eq-mot-00}
i D^{+}_{(t,r_{*})} F_{R} - \frac{ K  }{ C } \sigma^{2} F_{R}
-  ( \lambda \sigma^{i} a_{i} )^{+} \sigma^{1} F_{R} + i A^{-1}\partial_{r_{*}} \ln(\sqrt{A} C ) \sigma^{1} F_{R} 
+ i \frac{ 1 }{ 2 A C^2} \partial_{r_{*}} C^2 \sigma^{1} F_{R} - m F_{L} = 0, \\
%%%%%%%%%%%%%%%%%%%%%%%%%%%%%%%%%%%%%%%%%%%%%%%%5
\label{eq-mot-01}
i D^{-}_{(t,r_{*})} F_{L} + \frac{ K  }{ C } \sigma^{2} F_{L}
 -  ( \lambda \sigma^{i} a_{i} )^{-} \sigma^{1} F_{L} - i A^{-1} \partial_{r_{*}} \ln(\sqrt{A} C ) \sigma^{1} F_{L} 
- i  \frac{ 1 }{ 2 A C^2} \partial_{r_{*}} C^2 \sigma^{1} F_{L} - m  F_{R} = 0.
\een

%%%%%%%%%%%%%%%%%%%%%%%%%%%%%%%%%%%%%%%%%%%%%%%%%%%%%%%%%%%%%%%%%%%%%%%%%%%%%%%%%%%%%%%%%%%%%%%%%%%%
\subsection{ Massless Dirac fermions}
In this subsection we shall focus our attention on massless right-handed spinors.
One examines relation (\ref{eq-mot-00}), which in two-dimensional spacetime ($t,~r_{*}$)
may be rewritten as follows:
\be
i~ \sigma^{a}~ \nabla_{a} F_{R} -  \lambda~ \sigma^{a}~ B_{a} 
\sigma^{1} F_{R} - V~ \sigma^2~ F_{R} + 2~ i~ \sigma^{r_{*}}~ \partial_{r_{*}} \ln C~ F_{R} = 0,
\ee
where we set $V = \frac{ K }{ C }$ and $ B_{a} = e^{i}_{a} a_{i} $ being the 
{\it electric} part of Yang-Mills field. It can be readily find, by the direct computation, that the term $\sigma^{i} a_{i}$
now will be replaced by $\sigma^a B_{a}$.
Further, in order to get rid of $\partial_{r_{*}} \ln C$ factor one can rescale spinors in the following way:
\be
G_{R} \equiv  i \sigma^3 e^{ 2 \int \partial_{r_{*}} \ln C dr_{*}} F_{R}.
\ee
It will be useful to choose the new basis  for flat gamma matrices
\ben
\tilde{\gamma}^0 &=& - i \sigma^3, \qquad \tilde{\gamma}^1 = - \sigma^2, \\
\{ \tilde{\gamma}^a , \tilde{\gamma}^b \} &=& 2 \eta^{a b} ,\qquad \eta_{00} = -1 = - \eta_{11}, \\
\tilde{\gamma}^3 = \tilde{\gamma}^0 \tilde{\gamma}^1 &=& \sigma^1, \qquad 
\tilde{\gamma}_{L/R} = \frac { 1 }{ 2 }( I \pm \tilde{\gamma}^3 ).
\een 
On this account, equations of motion yield
\ben
\label{eq-GR}
i~ \tilde{\gamma}^{a}~ \nabla_{a} G_{R} +  \lambda~ \tilde{\gamma}^a~ B_{a}~ \tilde{\gamma}^3~ G_{R} 
- V~ \tilde{\gamma}^3~ G_{R} = 0.
\een
It can be deduced that they may be derived from the Lagrangian of the form
\be
\label{LR-1}
\mathcal{L}_{FR} = - i~ \bar{G}_{R}~ \tilde{\gamma}^{a}~ \nabla_{a} G_{R} - \lambda~  B_{a}~ 
\bar{G}_{R}~ \tilde{\gamma}^{a}~ \tilde{\gamma}^3~ G_{R} 
+ V~ \bar{G}_{R}~ \tilde{\gamma}_L{}~ G_{R} - V~ \bar{G}_{R}~ \tilde{\gamma}_{R}~ G_{R}.
\ee
It happened that $\mathcal{L}_{FR}$ can be examined by means of bosonization technique, i.e., the
fermionic degrees of freedom can be described by a scalar field propagating in ($t,~r_{*}$)-spacetime (see, e.g., \cite{bos})
One bosonizes the above Lagrangian by the following formulae
\ben
j^{a}_{R} \equiv  \bar{G}_{R}~ \tilde{\gamma}^{a}~ G_{R} &=& \frac{1}{ \sqrt{\pi}} 
\varepsilon^{a b}~ \nabla_{b} \phi_{R}, \qquad
j^{a}_{3 R} \equiv  \bar{G}_{R}~ \tilde{\gamma}^{a}~ \tilde{\gamma}^3~ G_{R} 
= \frac{ 1 }{ \sqrt{\pi} }~ \nabla^{a} \phi_{R}, \nonumber \\
\bar{G}_{R}~ \tilde{\gamma}_{L}~ G_{R} &=& b~ e^{2 i \sqrt{\pi} \phi_{R}}, \qquad  
\bar{G}_{R} \tilde{\gamma}_{R} G_{R} = b^{*}~ e^{- 2 i \sqrt{\pi} \phi_{R}}, 
\een 
where $b$ and  $b^{*}$ are constants depending on the normalization of the current 
$\bar{G} \gamma_{L/R} G$. For the brevity we set $b = b^{*}$. Thus the bosonized Lagrangian is provided by
\be
\mathcal{L}_{BR} = - \frac{1}{2} \nabla_{a}\phi_{R}~ \nabla^{a} \phi_{R} - \lambda~ B_{a}~ 
\frac{1}{\sqrt{\pi}} \nabla^a \phi_{R} +
V~ b~ ( e^{ 2 i \sqrt{\pi} \phi_{R}}  - e^{ - 2 i \sqrt{\pi} \phi_{R}} ), 
\ee  
while equation of motion for $\phi_{R}$ yields
\be
\label{phi_BR}
\nabla_{a} \nabla^{a} \phi_{R} +  \frac{ \lambda }{ \sqrt{\pi} } \nabla_{a} B^{a} 
+ 4~ i~ b~ \sqrt{\pi}~ V~ \cos( 2 \sqrt{\pi} \phi_{R}) = 0.
\ee

%%%%%%%%%%%%%%%%%%%%%%%%%%%%%%%%%%%%%%%%%%%%%%%%%%%%%%%%%%%%%%%%%%%%%%%%%%%%%%%%%%%%%%%%%%%%%%%%%%%%
Let us proceed to analyze the left-handed Dirac spinors.
One can use the following substitution in order to get rid of $\partial_{r_{*}} \ln C$ term
\be
G_{L} = i \sigma^3 e^{ 2 \int \partial_{r_{*}} \ln C dr_{*}} F_{L}.
\ee
We also choose the gamma matrices basis in the form as
\be
\tilde{\gamma}^0 = - i \sigma^3, \qquad \tilde{\gamma}^1 = + \sigma^2, \qquad
\tilde{\gamma}^3 = \tilde{\gamma}^0 \tilde{\gamma}^1 = - \sigma^1.
\ee
Having all the above in mind the equation of motion for 
$G_{L}$-spinors implies
\be
i~ \tilde{\gamma}^a~ \nabla_{a} G_{L} -  \lambda~ B_{a}~ \tilde{\gamma}^{a}~ \tilde{\gamma}^3~ G_{L} 
- V~ \tilde{\gamma}^3~ G_{L} = 0. 
\ee
On the other hand, the Lagrangian for $G_{L}$-spinors can be written as
\be
\label{LR-2}
\mathcal{L}_{FL} =  - i~ \bar{G}_{L}~ \tilde{\gamma}^{a}~ \nabla_{a} G_{L} + \lambda~  B_{a}~ \bar{G}_{L}~
 \tilde{\gamma}^{a}~ \tilde{\gamma}^3~ G_{L} 
+ V~ \bar{G}_{L}~ \tilde{\gamma}_{L}~ G_{L} - V~ \bar{G}_{L}~ \tilde{\gamma}_{R}~ G_{L}.
\ee
Replacing the fermionic degrees of freedom by the bosonization substitution given by the relations
\ben
j^{a}_{L} \equiv  \bar{G}_{L}~ \tilde{\gamma}^{a}~ G_{L} &=& 
\frac{1}{ \sqrt{\pi}} \varepsilon^{a b}~ \nabla_{b} \phi_{L}, \qquad
j^{a}_{3 L} \equiv  \bar{G}_{L}~ \tilde{\gamma}^{a}~ \tilde{\gamma}^3~ G_{L} 
= \frac{ 1 }{ \sqrt{\pi} } \nabla^{a} \phi_{L}, \nonumber \\
\bar{G}_{L}~ \tilde{\gamma}_{L}~ G_{L} &=& b~ e^{2 i \sqrt{\pi} \phi_{L}}, \qquad  
\bar{G}_{L}~ \tilde{\gamma}_{R}~ G_{L} = b^{*}~ e^{- 2 i \sqrt{\pi} \phi_{L}}, 
\een 
we achieve the bosonized Lagrangian for the scalar fields
\be
\mathcal{L}_{BL} = - \frac{1}{2}~ \nabla_{a}\phi_{L}~ \nabla^{a} \phi_{L} 
+ \lambda~ B_{a}~ \frac{1}{\sqrt{\pi}} \nabla^a \phi_{L} +
V~ b~ ( e^{ 2 i \sqrt{\pi} \phi_{L}}  - e^{ - 2 i \sqrt{\pi} \phi_{L}} ). 
\ee
Accordingly, equation of motion implies the following:
\be
\label{phi_BL}
\nabla_{a} \nabla^{a} \phi_{L}  
- \frac{ \lambda }{ \sqrt{\pi} } \nabla_{a} B^{a} + 4~ b~ i~\sqrt{\pi}~ V~ \cos( 2 \sqrt{\pi} \phi_{L} ) = 0.
\ee
One can remark that the only difference between relations for $\phi_{L}$ and $\phi_{R}$ is the sign
of the term containing $B_{a}$. 
However, in Ref.\cite{biz92} it was pointed out that in order
to have  finite Yang-Mills black hole mass one needed to get $B_{a} = 0$. By virtue of this argument
we can readily verify that equations of motion for {\it right} and {\it left-handed } fermion fields are identical.
Therefore, in what follows, we restrict our attention to only one equation of motion.
\par
Having in mind the arguments quoted above, we commence with the asymptotic behaviour 
analysis of field $\phi_{R}$. From the point of view of demanding asymptotical flatness
of the black hole solution in question, one has that
$g_{tt} \sim g_{rr} \sim 1$,
$r \sim r_{*}$ and $C = r$, as $r$-coordinate tends to infinity.
It enables us to write the underlying equation of motion in the form as
\be
- \partial_{t}^2 \phi_{R} + \partial^2_{r_{*}} \phi_{R}  +
4~ i~ b~ \sqrt{\pi}~ \frac{K(\infty)} {r_{*}}~ \cos( 2 \sqrt{\pi} \phi_{R}) = 0, 
\label{asym}
\ee
where we set $V \sim \frac{ K(\infty) }{ r_{*} }$, $K(\infty) = \pm 1$.
In order to satisfy our demands about finiteness
of $\phi_{R}$ as $r_{*}$ tends to infinity
the last term 
in relation (\ref{asym}) vanishes. In terms of it, we arrive at the equation 
\be
- \partial_{t}^2 \phi_{R} + \partial_{r_{*}}^2 \phi = 0,
\label{asymp1}
\ee 
with the regular solution provided by
\be
\phi_{R} = d~ e^{ - i \omega ( t \pm r_{*})},
\ee 
where $d$ is an arbitrary constant.
The obtained solution is time-dependent and admits the non-zero
fermion current at infinity, i.e., $\na_j\phi_{R} \neq 0$. 
\par
Let us refine our studies to the near-horizon geometry of YM black hole surrounded by
the Dirac fermion fields. In the vicinity of the event horizon one achieves
\be
A^{2} = N^2 e^{\delta(r_{h})}, \qquad A^{2} = N^2 e^{\delta(r_{h})}, \qquad N^2 = 2 \kappa ( r - r_{h}),
\ee
where $\kappa$ is the surface gravity and $e^{\delta(r_{h})}$ is a constant factor.\\
Taking the usual change of variables $ r - r_{h} = \rho^{-1}$, we obtain that
\be
r_{*} = - c~ \rho,
\ee
where $c = \frac{ 1 }{ 2 \kappa e^{ \sqrt{ \delta(r_{h}) }}} $. Exploring
these relations we can rewrite equation of motion in the form as
\be
- \partial_{t}^2 \phi_{R} + c^2~ \partial_{\rho}^2 \phi_{R}  + 
 i~ \frac{ 8~ b~ \sqrt{\pi}~ K(r_{h})~ \kappa~ e^{ \sqrt{\delta(r_{h})} } }{ r_{h}~ \rho}
~ \cos( 2 \sqrt{\pi} \phi_{R} ) = 0.
\ee
By the same reasoning as we followed in deriving equation (\ref{asymp1}), we conclude
that dropping the last term in the above relation, the solution implies
\be
\phi_{R} = c_1~ e^{ - i \omega ( t \pm \frac{ 1 }{ c^2 } \rho ) },
\ee
where we have set $c_1$ as integration constant.\\
Summing it all up, we conclude that the asymptotic analysis
of scalar field equations being bosonized Dirac fermions, conducts to the time-dependent
plane wave solutions. This is in contradiction to the static nature of the considered
YM black hole. On this account, it is impossible to obtain static spherically symmetric
YM black hole solution surrounded by {\it Dirac fermion vacuum}.

%%%%%%%%%%%%%%%%%%%%%%%%%%%%%%%%%%%%%%%%%%%%%%%%%%%%%%%%%%%%%%%%%%%%%%%%%%%%%%%%%%%%%%%%%%%%%%%%%%%%%%%%%%%%%%%%%%%%%%%%%%%%%%%%%%%%%%%%%%%%%%%%%%%%%%%%%%%%%%%%%%%%%
\subsection{Massive Dirac fermions.}
In this subsection we are mainly concerned with the Dirac massive case. In what follows
we take into account the following ansatze for
$F_{L}$ and $F_{R}$:
\ben
F_{L} &=& i \sigma^3 F_{R} \equiv e^{ - \int 2 \partial_{r_{*}} \ln C dr_{*}} G \nonumber \\
F_{R} &=& - i \sigma^3 e^{ - \int 2 \partial_{r_{*}} \ln C dr_{*}} G.
\een
Because of the fact that 
the left-handed fermions can be expressed in terms of a linear combination of the right-hand ones, we will 
use only the right-hand part of the original fermionic Lagrangian supplemented by the appropriate mass 
term. Namely, we obtain the equations of the form as   
\be
i~ \sigma^a~ \nabla_{a} F_{R} - \frac{ K }{ C }~ \sigma^{2}~ F_{R}
-  \lambda~ \sigma^{a}~ B_{a}~ \sigma^{1}~ F_{R} + 2~ i~ A^{-1}~\partial_{r_{*}} 
\ln( C )~ \sigma^{1}~ F_{R} - m~ i~ \sigma^3~ F_{R} = 0. 
\ee
On this account it is customary to write 
the following relations for $G$-fermions:
\be
i~ \tilde{\gamma}^a~ \nabla_{a} G - V~ \tilde{\gamma}^3~ G + 
\lambda~ \tilde{\gamma}^a~ B_{a}~ \tilde{\gamma}^3~ G - m~ G = 0,
\ee
where 
$\tilde{\gamma}$ 
are gamma matrices written in 
basis which is chosen like in the massless right-handed case. We also have that $V = \frac{ K }{ C }$.
Hence,
the effective Lagrangian for G-fermions will be provided by the expression
\be
\mathcal{L}_{G F} = - i~ \bar{G}~ \tilde{\gamma}^a~ \nabla_{a} G 
-  \lambda~ B_{a}~  \bar{G}~ \tilde{\gamma}^a~ \tilde{\gamma}^3~ G +
( V + m )~ \bar{G}~ \tilde{\gamma}_{L}~ G + ( m - V)~ \bar{G}~ \tilde{\gamma}_{R}~ G. 
\ee 
Next, as in the preceding sections, we try with the bosonization scheme
\ben
j^{a} \equiv  \bar{G}~ \tilde{\gamma}^{a}~ G &=& \frac{1}{ \sqrt{\pi}} \varepsilon^{a b}~ \nabla_{b} \phi, 
\qquad
j^{a}_{3} \equiv  \bar{G}~ \tilde{\gamma}^{a}~ \tilde{\gamma}^3~ G = \frac{ 1 }{ \sqrt{\pi} } \nabla^{a} \phi, 
\nonumber \\
\bar{G}~ \tilde{\gamma}_{L}~ G &=& b~ e^{2 i \sqrt{\pi} \phi }, \qquad  
\bar{G}~ \tilde{\gamma}_{R}~ G = b^{*}~ e^{- 2 i \sqrt{\pi} \phi}. 
\een 
When we set $b = b^* $, one enables to find  
Lagrangian for scalar field given by 
\be
\mathcal{L}_{G B} = - \frac{1}{2} \nabla_{a} \phi~ \nabla^a \phi - \frac{ \lambda }{ \sqrt{\pi} }~ B_{a}~
 \nabla^a \phi +
( V + m )~ b~ e^{ 2 i \sqrt{\pi} \phi} + ( m - V )~ b~ e^{ - 2 i \sqrt{\pi} \phi},
\ee
where $b$ is constant. On the other hand, equation of motion for $\phi$ field implies
\be
\nabla_{a} \nabla^a \phi + \frac{ \lambda }{ \sqrt{\pi} } \nabla_{a} B^a + 
2~ i~ b~ \sqrt{\pi} ~
\bigg \{ 
V~ [ e^{ 2 i \sqrt{\pi} \phi} + e^{ - 2 i \sqrt{\pi} \phi} ] + 
m~ [ e^{ 2 i \sqrt{\pi} \phi }  - e^{ - 2 i \sqrt{\pi} \phi} ]
\bigg \} = 0, 
\ee
which can be rewritten in a more compact form
\be
\nabla_{a} \nabla^a \phi + \frac{ \lambda }{ \sqrt{\pi} } \nabla_{a} B^a + 
4~ i~ b~\sqrt{\pi} ~
\bigg \{ 
V~ \cos( 2 \sqrt{\pi} \phi ) + i~ m~ \sin( 2 \sqrt{\pi} \phi)
\bigg \} = 0. 
\ee
Applying the same analysis as in the preceding section lead us to the conclusion that in the
near-horizon limit the bosonization field $\phi$ will be given in the form of a plane wave.
Namely, it will be described by
\be
\phi = d_1~ e^{ - i \omega ( t \pm \frac{ 1 }{ c^2 } \rho ) },
\ee
where $d_1$ is integration constant while $c$ is the same constant like in massless case.\\
On the other hand, in the limit when $r_{*} \rightarrow \infty$ the underlying
equation of motion may be written as 
\be
- \partial_{t}^2 \phi + \partial^2_{r_{*}} \phi - 
4~ b~ \sqrt{\pi}~ m~ \sin( 2 \sqrt{\pi} \phi) = 0, 
\ee
where due to the previously quoted arguments we have omitted  $B_{a}$. Because of the fact
that $V \sim 1/r$, the term with cosine function tends to zero. Just we attain the form
of the well known sine-Gordon equation, the solution of which implies
\be
\phi = \frac{ 2 }{ \sqrt{\pi} }~ \arctan \bigg( e^{ - \sqrt{ \frac{ 8 b \pi m }{ 1 - v^2 }  }( r_{*} - vt )  } 
\bigg),
\ee
where $v$ is integration constant. From the above relation it can be inferred that fermion
current tends to zero as $r_{*} \rightarrow \infty$.\\
To conclude this section, we remark that if we put $B_{a} = 0$ we obtain
the time-dependent Dirac fermions as well as through equations of motion, the time-dependent
YM fields and the coefficients of the studied line element. All the above contradicts our primary assumptions
about staticity of the the system in question.

%%%%%%%%%%%%%%%%%%%%%%%%%%%%%%%%%%%%%%%%%%%%%%%%%%%%%%%%%%%%%%%%%%%%%%%%%%%%%%%%%%%%%%%%%%%%%%%%%%%%
%%%%%%%%%%%%%%%%%%%%%%%%%%%%%%%%%%%%%%%%%%%%%%%%%%%%%%%%%%%%%%%%%%%%%%%%%%%%%%%%%%%%%%%%%%%%%%%%%%%%
\section{Backreaction on Yang-Mills fields.}

If one considers a spherically symmetric spacetime being a Lorentz manifold on which
$SO(3)$ group acts like isometry in such way that all group orbits are $S^2$-spheres,
the spacetme in question may be locally written as a warped product of a two-dimensional
Lorentz manifold and two-sphere with the standard metric on it \cite{vol99}-\cite{cor76}.
It happened that
it is convenient to rederive the field equations in the spherically symmetric case by varying the
effective two-dimensional action. Putting in EYM action the ansatz for the YM gauge fields and
the line element describing the symmetry in question, one can obtain the two-dimensional Lagrangian
which yields
\be
\mathcal{L}_{YM} = - \frac{C^2}{4}~ f_{ab}f^{ab} - |d K|^2 - \frac{1}{ 2 C^2}~ ( |K|^2 - 1)^2,
\ee
where the {\it covariant derivative} and the strength of $B_{a}$ are given by the relations
\be
d = \na_{a} - i~ B_{a}, \qquad f_{ab} = \na_{a} B_{b} - \na_{b} B_{a}.
\ee
We have denoted by a subscript $a =t,~r_{*}$-coordinates, respectively.
Equations of motion in the presence of the bosonized massless fermions are provided by
\ben \label{ym1}
\nabla_{a} [ C^2 f^{a b} ] &-& 2~ |K|^2~ B^{b} - \lambda~ j^{b}_{3 R} + \lambda ~j^{b}_{3 L} = 0,\\ \label{ym2}
\nabla_{a} \nabla^{a} K &-& \frac{ 2 }{ C^2 }~ K~ ( |K|^2 - 1 ) + 2~ K~ B_{a}B^{a} +
\frac{ b }{ C }~ [ e^{ 2 i \sqrt{\pi} \phi_{R} } - e^{- 2 i \sqrt{\pi} \phi_{R} }  
+ e^{ 2 i \sqrt{\pi} \phi_{L} } - e^{ - 2 i \sqrt{\pi} \phi_{L} }  ] = 0, 
\een   
where $j^{a}_{3 R} = \bar{G}_{R}~ \tilde{\gamma}^{a}~ \tilde{\gamma}^3~ G_{R} $ 
and $j^{a}_{3 L} = \bar{G}_{L}~ \tilde{\gamma}^{a}~ \tilde{\gamma}^3~ G_{L} $. 
They can be reduced to the forms as follows:
\ben
\label{f_ab-massles}
\nabla_{a} [ C^2 f^{a b} ] &-& 2~ |K|^2~ B^{b} 
- \frac{ \lambda } { \sqrt{\pi} }~[ \nabla^{b} \phi_{R} - \nabla^{b} \phi_{L}  ] = 0, \\
%%%%%%%%%%
\label{K-massles}
\nabla_{a} \nabla^{a} K &-& \frac{ 2 }{ C^2 }~ K~ ( |K|^2 - 1 ) + 2~ K~ B_{a}B^{a} - 
i~ \frac{ 2 b }{ C }~ [ \sin( 2 \sqrt{\pi} \phi_{R}) + \sin(2 \sqrt{\pi} \phi_{L} ) ] = 0.
\een
Having in mind equation (\ref{phi_BR}) and (\ref{phi_BL}), in the case when $B_{a} = 0$, one can see that
$j^{b}_{3 R} = j^{b}_{3 L}$.
Thus, $B_{a} = 0$ is the trivial solution of (\ref{ym1}). As far as the influence fermions on YM fields
is concerned, from relation (\ref{K-massles}) it can be inferred  that the last term is not equal to zero and 
this leads to the conclusion that fermions have influenced on the {\it magnetic} part of YM fields.
By virtue of the above, we observe that there is nonzero fermion contribution to 
{\it magnetic} part of Yang-Mills field. Since 
$\phi_{L}$ and $\phi_{R}$ are time dependent,
then {\it magnetic} part of YM fields $K$ has also time dependence, which in turn cause that the 
coefficients of the line element in question also count on time.\\
%%%%%%%%%%%%%%%%%%%%%%%%%%%%%%%%%%%%%
In the case when $B_{a} \neq 0$, from equations (\ref{phi_BR}) and (\ref{phi_BL}), one can find  
fermion currents $j^{b}_{3 R},~j^{b}_{3 L}$. On the other hand, the inspection of equation (\ref{ym1}) reveals the 
fact that fermion current causes the nontrivial solution for $B_{a}$. Consequently,
black hole in question has both {\it magnetic} and {\it electric} charges. One gets the dyonic black hole.
But this in turn leads to the infiniteness of the mass of the considered object \cite{biz92}.
\par
As was remarked, taking into account massive fermions, one has that $F_{L} = i \sigma^3 F_{R}$. This fact reduces the
number of the independent fields, i.e., instead of $\phi_{L}$ and $\phi_{R}$ we have only one 
field $\phi$, satisfying the relation
\be
\label{f_ab-massive}
\nabla_{a} \bigg( C^2 f^{a b} \bigg) - 2~ |K|^2~ B^{b} - \frac{ \lambda } { \sqrt{\pi} } \nabla^{b} \phi  = 0. 
\ee
Hence, we get the black hole with both electric and magnetic charges (the dyon). On the other hand, it happened
that this dyonic black hole can not have finite mass \cite{biz92}.
\par
The above analysis reveals the fact that the presence of the Dirac fermion field leads to the destruction of a static 
YM black hole solution. First of all, the scalar which we get in the bosonization process are time-dependent. This in 
turn causes that {\it magnetic} part of YM fields as well as the coefficients of the line element in question
depends also on time. It destroys are assumption about staticity of the considered black hole. These conclusions are
true both for massless and massive Dirac fermions. Secondly, we can readily see that the presence of the Dirac
fermion fields will provide the existence of the {\it electric} part of the YM fields, which in turn leads to the infiniteness
of black hole mass.

%%%%%%%%%%%%%%%%%%%%%%%%%%%%%%%%%%%%%%%%%%%%%%%%%%%%%%%%%%%%%%%%%%%%%%%%%%%%%%%%%%%%%%%%%%%%%%%%%%%%
%%%%%%%%%%%%%%%%%%%%%%%%%%%%%%%%%%%%%%%%%%%%%%%%%%%%%%%%%%%%%%%%%%%%%%%%%%%%%%%%%%%%%%%%%%%%%%%%%%%%
\section{ Five-dimensional Yang-Mills black hole.}
It happens that studies of spherically symmetric solutions in higher dimensional theories reveals two ways of researches.
One is connected with the assumption that we have spherically symmetric solutions in $n$-dimensions 
(this attitude is important in high-energy problems)
and the other, when one assumes that solutions are spherically symmetric only in four-dimensional manifold. 
The second approach is importance from the point of view of the present Universe. These idea
were explored in Refs.\cite{plus}, where $(4+1)$-dimensional EYM systems were elaborated. It turned
out that EYM system reduces to an effective four-dimensional EYMHd model. The $(4+n)$-dimensional case was considered 
in Ref.\cite{bri04},
with the assumption that all of the $n$-dimensional fields are independent on the extra dimensions. The solutions were 
spherically symmetric in four-dimensions while the additional dimensions were bounded with Ricci flat manifold.
\par
Now, we shall proceed to elaborate $(4+1)$-dimensional EYM theory, where both matter fields and
line element coefficients are independent on the fifth coordinate. Let us suppose that five-dimensional metric and
five dimensional field are parameterized as follows:
\ben
{}^{(5)}ds^2 = g_{M N}~dx^N~dx^M &=& e^{ - \xi }[ - A^2 dt^2 + B^2 dr^2 + C^2 d \Omega^2 ] + e^{2 \xi} (d x^5)^2,\\
H_{M}^{a} dx^M &=& H_{\mu}^{a} dx^{\mu} + \Phi^{a} dx^5,
\een
where $M,~N = t,~ r,~ \theta,~\phi,~ x^5$, $a$ is a group index, $H_{M}$ is the five-dimensional Yang-Mills field.
On the other hand, $H_{\mu}$ denotes the four-dimensional Yang-Mills field components and
$\xi$ plays the role of the dilaton \cite{plus}. The above relations provide that we attain to the following effective
four-dimensional EYMD theory for which the Lagrangian after compactification yields
\be
\mathcal{L}_{4} =  a_{1} R - a_{2} \nabla_{\mu} \xi \nabla^{\mu} \xi 
- \frac{ 1 }{ 4 } e^{ \xi } F_{\mu \nu}^{a} F^{a \mu \nu} - \frac{ 1 }{ 2 }e^{ - 2 \xi} 
{ \it D}_{\mu} \Phi^a { \it D}^{\mu} \Phi^a,
\ee 
where $a_{1}$ and $a_{2}$ are constants depending on five dimensional gravitational constant 
while the {\it covariant derivative} of the Higgs field in the adjoint representation implies
\be
D_{\mu} \Phi^a = \nabla_{\mu} \Phi^a + \varepsilon_{a b c} H^{b}_{\mu} \Phi^{c}.
\ee
Let us assume further, that $\Phi = \nu Y(r) \tau^i n_{i}$, where 
$\nu$ is the expectation value of the Higgs field.\\
The matter Lagrangian written in $(t, r_{*})$-coordinates is given by
\be
\mathcal{L}_{2-dim} =  - a_{2}~ C^2~ \nabla_{a} \xi \nabla^{a} \xi 
- \frac{ C^2 }{ 4 }~ e^{ \xi }~ f_{a b} f^{a b} - e^{ \xi }~ (d K)^2 - \frac{ 1 }{ 2 C^2 }~e^{ \xi }~( |K|^2 - 1)^2
- \frac{ C^2 }{ 2 }~e^{ - 2 \xi}~ \nabla_{a} Y \nabla^a Y - e^{ - 2 \xi }~ |K|^2~ Y^2.
\ee
As far as the five-dimensional fermions is concerned, after dimensional reduction their action yields
\be
S_{F_4} = \int \sqrt{-{}^{(4)}g}~  d^4 x~ [ i~ \bar{\psi}~ 
\gamma^{\mu}~ \stackrel{\leftrightarrow}{D}_{\mu} \psi 
+ \lambda_{2}~ \bar{ \psi}~ \gamma^{5}~ e^{ -2 \xi }~ \Phi^{i}~ \tau^{i}~ \psi ], 
\ee
where  $\gamma^5 = \pmatrix{ I& 0 \cr 0 &-I  } $. 
One can observe that they gain mass by coupling to the five-dimensional component of YM field $\Phi^{i}$. 
Namely it implies
\be
- m~ [ \bar{\psi}_{R}~ \psi_{L} + \bar{\psi}_{L}~ \psi_{R}  ] \rightarrow 
\lambda_{2}~  e^{ - 2 \xi}~  \bar{\psi}_{R}~ \Phi^i~ \tau^{i}~ \psi_{L}
- \lambda_{2}~ e^{ - 2 \xi}~  \bar{\psi}_{L}~ \Phi^i~ \tau^i~  \psi_{R},
\ee
where $\lambda_{2}$ is fermion coupling constant and $\tau^i$ are $SU(2)$ generators.
On the other hand, $\Phi^{i}$ resembles Higgs fields in four-dimensional background.
Massless fermions will be described by the same equations and therefore we restrict our attention to
the massive case.\\
Further, we represent $\psi_{L/R}$ in terms of $\chi_{1/2}$ 
\be
\lambda_{2}~ e^{ - 2 \xi}~ \bar{ \psi}_{R}~ \Phi^{i}~ \tau^{i}~ \psi_{L} 
= - C^2~ \lambda_{2}~ \nu~ Y~ e^{ - 2 \xi }~ \bar{F}_{R}~ \sigma^1~ F_{L}.
\ee
Expressing fermions as two-dimensional ones and integrating over
the angles, we get the effective two-dimensional Lagrangian
given by (\ref{psi-2dim}) with following substitutions:
\ben
 - m~ \bar{F}_{R}~ F_{L} &\rightarrow& - \lambda_{2}~ \nu~ e^{ - 2 \xi }~ Y~ \bar{F}_{R}~ \sigma^1~ F_{L}, \nonumber \\ 
 - m~ \bar{F}_{L}~ F_{R} &\rightarrow&  \lambda_{2}~ \nu~ e^{ - 2 \xi }~ Y~ \bar{F}_{L}~ \sigma^1~ F_{R}.
\een
Bosonization of the Dirac fermion fields will take place as in  the preceding
sections. Namely, we introduce $G$-fermions $F_{L} = i~ \sigma^3~ F_{R}$ and
$F_{R} = - i~ \sigma^3~ e^{ - \int 2 \partial_{r_{*}} \ln( C ) d r_{*}}~ G$, where
two-dimensional base of the gamma matrices is chosen as follows:
\be
\tilde{\gamma}^0 = -i~ \sigma^3, \qquad \tilde{\gamma}^1 = - \sigma^2, \qquad
\tilde{\gamma}^3 = \tilde{\gamma}^0 \tilde{\gamma}^1.  
\ee
It leads us to the equation of motion of the form as
\be
i~ \tilde{\gamma}^a~ \nabla_{a}~ G - V^{'}~ \tilde{\gamma}^3~ G + \lambda~ \tilde{\gamma}^a~ B_{a}~ 
\tilde{\gamma}^3~ G = 0,
\ee
where we set $V^{'} = \frac{ K }{ C } + \lambda_{2}~ \nu~  e^{ - 2 \xi }~ Y $. They can be derived
from the effective Lagrangian provided by
\be
\mathcal{L}_{GF} = - i~ \bar{G}~ \tilde{\gamma}^a~ \nabla_{a}~ G 
- \lambda~ \bar{G}~ \tilde{\gamma}^a~ B_{a}~ \tilde{\gamma}^3~ G +
V^{'}~ \bar{G}~ \tilde{\gamma}_{L}~ G - V^{'}~ \bar{G}~ \tilde{\gamma}_{R}~ G.
\ee
It can be easily verified that using following bosonization formulae
\ben
j^a = \bar{G}~ \tilde{\gamma}^a~ G &=& \frac{ 1 }{ \sqrt{\pi} }~ \varepsilon^{a b}~ \nabla_{b} \phi, \qquad
j^{a}_{3} = \bar{G}~ \tilde{\gamma}^a~ \tilde{\gamma}^3~ G = \frac{ 1 }{ \sqrt{\pi} }~ \nabla^a \phi, \nonumber \\
\bar{G}~ \tilde{\gamma}_{L}~ G &=& b~ e^{2 i \sqrt{\pi}~ \phi }, \qquad  
\bar{G}~ \tilde{\gamma}_{R}~ G = b^{*}~ e^{- 2~ i~ \sqrt{\pi}~ \phi }, 
\een 
and setting $b = b^*$, the bosonized Lagrangian becomes
\be
\mathcal{L}_{GB} = - \frac{ 1 }{ 2 }~ \nabla_{a} \phi \nabla^a \phi - \frac{ \lambda }{ \sqrt{\pi} }~ B_{a}~
 \nabla^a \phi 
+ V^{'}~ b~ [ e^{2 i \sqrt{\pi} \phi } - e^{- 2 i \sqrt{\pi} \phi } ]. 
\ee
Consequently, equation of motion for scalar field $\phi$ is given by
\be
\nabla_{a} \nabla^a \phi + \frac{ \lambda }{ \sqrt{\pi} }~ \nabla_{a} B^a + 
2~ i~ b~ \sqrt{\pi}~ V^{'}~ [ e^{2 i \sqrt{\pi} \phi } + e^{- 2 i \sqrt{\pi} \phi } ] = 0,
\ee
or in a more compact form it implies
\be
\nabla_{a} \nabla^a \phi + \frac{ \lambda }{ \sqrt{\pi} }~ \nabla_{a} B^a + 
4~ i~ b~ \sqrt{\pi}~ V^{'}~ \cos( 2 \sqrt{\pi} \phi ) = 0.
\ee

%%%%%%%%%%%%%%%%%%%%%%%%%%%%%%%%%%%%%%%%%%%%%%%%%%%%%%%%%%%%%%%%%%%%%%%%%%%%%%%%%%%%%%%%%%%%%%%%%%%%
\subsubsection{ Asymptotic analysis of equations of motions. }
Taking the same change of variable like in massless case, namely $ r_{*} \sim - \rho$, and 
noticing that $A^2 \sim \rho^{-1}$
we see that in near-horizon limit $\phi$ is given by the plane wave solution.
For the finiteness of black hole mass we assume that $B_{a} = 0$, then one arrives at
\be
\phi = e_1~ e^{ - i \omega( t \pm \frac{1}{c^2} \rho  )  },
\ee
where $e_1$ is arbitrary constant while $c^2$ is the same like in 
the previously studied massless case. Thus, 
if we demand $B_{a} = 0$, the  regular solution for $\phi$ must be time-dependent.
\par
Now we will look for solution in $r_{*} \rightarrow \infty $ limit. Namely, one has that
\be
- \partial_{t}^2 \phi + \partial_{r_{*}}^2 \phi + 4~ i~ b~ \sqrt{\pi}~ m_{0}~ \cos( 2 \sqrt{\pi} \phi) = 0,
\ee
where $m_{0} = \lambda_{2}~ \nu~ Y_{0}~ e^{ - 2 \xi_{0}} $, $Y_{0}$ and $\xi_{0}$  are 
the asymptotic values of Higgs and dilaton fields, respectively.
One can rewriting $\phi$ as 
$\phi(r_{*}) = \phi_{a}(r_{*}) - \frac{ 1 }{ 2 \sqrt{\pi} } \frac{ \pi }{ 2 }  $, which
leads us to the following form of equations of motion 
\be
- \partial_{t}^2 \phi_{a} + \partial_{r_{*}}^2 \phi_{a} - 4~ i~ b~ \sqrt{\pi}~ m_{0}~ \sin( 2 \sqrt{\pi} \phi_{a}) = 0.
\ee
As we see it is sine-Gordon equation with complex coefficients. Its solution yields
\be
\phi_{a} = \frac{ 2 }{ \sqrt{\pi} } \arctan \bigg( e^{- \sqrt{ \frac{ 8 i \pi b m_{0} }{ 1 - v^2}  }( r_{*} - vt )  } 
\bigg),
\ee
where $\nu$ is an integration constant.\\
It can be remarked that $\phi$ tends to
the constant value $- \frac{ \sqrt{\pi} }{ 4 } $ as $r_{*} $ goes to infinity. But $\phi$ itself is not
a physical quantity. Physical quantities are fermions fields and they are given by derivatives of $\phi$. 
Concluding, in both cases of massive fermions (normal and Higgs generated mass) 
fermionic currents decay at infinity.\\  
The analysis of fermion backreaction on Yang-Mills field goes along the same line like in pure four-dimensional case.
Conclusion are qualitatively the same. Namely, time-dependent fermion field leads to the destruction of static ansatz for 
Yang-Mills black hole (massive and massless case), or massive fermions lead to the appearance of the nonzero 
electric part of YM field.

%%%%%%%%%%%%%%%%%%%%%%%%%%%%%%%%%%%%%%%%%%%%%%%%%%%%%%%%%%%%%%%%%%%%%%%%%%%%%%%%%%%%%%%%%%%%%%%%%%%%
%%%%%%%%%%%%%%%%%%%%%%%%%%%%%%%%%%%%%%%%%%%%%%%%%%%%%%%%%%%%%%%%%%%%%%%%%%%%%%%%%%%%%%%%%%%%%%%%%%%%
\section{Conclusions}
In our paper we have considered the influence of the Dirac fermion field on EYM black hole. One takes into account
two cases, i.e., the four-dimensional YM black hole \cite{biz92} and black hole in five-dimensional
generalization of YM theory. In the latter case the five-dimensional theory reduces to the four-dimensional 
EYMHd model. In both cases we elaborated $SU(2)$ YM theory and treated Dirac fermion in s-wave sector.
Assuming a spherically symmetric static asymptotically flat black hole spacetime, we bosonize fermion fields and 
study equations of motion for the obtained scalar fields. 
\par
In a massless fermion sector we arrive
at two scalar fields $\phi_{L}$ and $\phi_{R}$ corresponding respectively to the left and right-handed Dirac fermions.
It happened that the action governing the scalar fields differs only by the sign in the term connected with
{\it electric} part of YM field. Because of the fact that the finite mass YM black hole configuration exists only when
{\it electric} part of YM field is equal to zero, the resulting equations of motion for both scalar fields are identical.
The analysis of $\phi_{L/R}$ fields in the near-horizon and near infinity limits, reveals the fact that they are given 
by plane wave solutions. It in turns leads to the time-dependence of {\it magnetic} part of YM field as well as through
EYM field equations to the time-dependence of the considered line element coefficients.
\par
The situation is slightly different in massive fermions case. First of all one should recognize two kinds of 
mass terms emerging in our considerations. Namely, the {\it ordinary} mass term $m \bar{\psi} \psi$
appears in four-dimensional spacetime and the mass term connected with Higgs field 
$\lambda \Phi \bar{\psi} \psi$ in five-dimensional case which reduces effectively to the four-dimensional
EYMHd theory.
Before bosonization we use simplifying assumption that the right and left-handed part of the Dirac fermion field
are connected through the transformation $F_{R} = i \sigma^3 F_{L}$. It allows us to express Dirac fermion field by 
only one
scalar field $\phi$. Next step was to analyze the beahaviour of  $\phi$ filed in the near-horizon and infinity limits.
It turns out that in the near-horizon limit the solution is described by a plane wave. This conclusion is true for both
aforementioned masses. On the other hand, as far as the near infinity limit is concerned, equations of motion 
for {\it ordinary} mass term $m \bar{\psi} \psi$ reduces to the sine-Gordon type of equation.
This equation has a time-dependent decaying solution as one approaches near infinity limit (the so-called {\it anti-kink}
solution). In the case of the {\it Higgs generated} mass term $\lambda \Phi \bar{\psi} \psi$, equations in question 
can be also brought to the sine-Gordon type of equation but with a complex coefficient. It can be found that the solution
reduces to the decaying oscillation function plus a nonzero constant term.
Moreover, the presence of the Dirac fermion field currents will cause non-trivial value of the {\it electric} 
part of the YM fields, which in turn leads to the infiniteness
of black hole mass.
\par
Summing it all up, we remark that the presence of Dirac fermion field ({\it fermion vacuum}) in  the spacetime
of magnetically charged spherically symmetric static asymptotically flat YM black hole will lead to the destruction
of this black object. It will happen by the destroying the static ansatz for black hole by both massive 
and massless Dirac fermions. Asymptotical analysis of the behaviour of the fermion fields suggests that the 
massless fermions are propagating in the whole spacetime, while massive Dirac fermions are confined to the
near-horizon region. This conclusion remains true for the effectively reduced five-dimensional YM theory.
Just, in order to have static YM black hole one should have Dirac fermions which are constant in the domain of 
outer communication of the black hole in question or Dirac fermion fields ought to enter the black hole (massive
Dirac field) or escape to infinity (massless Dirac fermions). This is just the same conclusion as 
conceived in Refs.\cite{smo}.

%Second possibility of destruction of black hole is through appearance of nonzero electric Yang-Mills filed ($B_{a}$).
%This could happen if fermion currents give nonzero contribution to equation of motion for $f_{ab}$. 
%As turns out this wasn't the case for massless fermions. To see this we note that initially $B_{a} = 0$ and equations
%of motion for $\phi_{L/R}$ are identical. Also we note that full equations differ only by sign of coupling to $B_{a}$.
%Having this in mind we find that contribution of $\phi_{L}$ and $\phi_{R}$ cancel each other. So there is no
%net contribution to equation for $B_{a}$ and $B_{a} = 0$ is still correct solution. 

%%%%%%%%%%%%%%%%%%%%%%%%%%%%%%%%%%%%%%%%%%%%%%%%%%%%%%%%%%%%%%%%%%%%%%%%%%%%%%%%%%%%%%%%%%%%%%%%%%%%%%%
%%%%%%%%%%%%%%%%%%%%%%%%%%%%%%%%%%%%%%%%%%%%%%%%%%%%%%%%%%%%%%%%%%%%%%%%%%%%%%%%%%%%%%%%%%%%%%%%
%\begin{appendix}

%\section{Irred   } 
%\label{irtf}
%\end{appendix}
\begin{acknowledgments}
{\L} N was supported by Human Capital Programme of European Social Fund sponsored
by European Union.\\
MR was partially supported by the grant of the National Science Center
$2011/01/B/ST2/00408$.
\end{acknowledgments}
%%%%%%%%%%%%%%%%%%%%%%%%%%%%%%%%%%%%%%%%%%%%%%%%%%%%%%%%%%%%%%%%%%%%%%%%%%%%%%%%%%%%%%%%%%
%%%%%%%%%%%%%%%%%%%%%%%%%%%%%%%%%%%%%%%%%%%%%%%%%%%%%%%%%%%%%%%%%%%%%%%%%%%%%%%%%%%%%%%%%%%%%%%%%%%%%%%
%%%%%%%%%%%%%%%%%%%%%%%%%%%%%%%%%%%%%%%%%%%%%%%%%%%%%%%%%%%%%%%%%%%
%%%%%%%%%%%%%%%%%%%%%%%%%%%%%%%%%%%%%%%%%%%%%%%%%%%%%%%%%%%%%%%%%%%%%%%%%%%%%%%%%
%%%%%%%%%%%%%%%%%%%%%%%%%%%%%%%%%%%%%%%%%%%%%%%%%%%%%%%%%%%%%%%%%%%%%%%%%%%%%%%%%

%%%%%%%%%%%%%%%%%%%%%%%%%%%%%%%%%%%%%%%%%%%%%%%%%%%%%%%%%%%%%%%%%%%%%%%%%%%%%%%%%%%%%%%%%%%%%%%%%%%%%%%%%%%%%%
%%%%%%%%%%%%%%%%%%%%%%%%%%%%%%%%%%%%%%%%%%%%%%%%%%%%%%%%%%%%%%%%%%%%%%%%%%%%%%%%%%%%%%%%%%%%%%%%%%%%%%%%%%%%%

%\begin{figure}[p]
%\includegraphics[scale=0.5,angle=270]{fig12.eps}
%\caption{Dependence of Plot of $|\xi_{i}|^2$ where $\xi_{i} = \{ \psi_{L},~\chi_{L} \}$ on fermion mass for 
%extremal black string. 
%Rest of the parameters are $N = 1$, $q_{r} = 5$, $q_{e} = 10$, $\omega =k = 0$.}
%\label{fig12}
%\end{figure}

%%%%%%%%%%%%%%%%%%%%%%%%%%%%%%%%%%%%%%%%%%%%%%%%%%%%%%%%%%%%%%%%%%%%%%%%%%%%%%%%%%%%%%%%%%%%%%%%%%%%%%%%%%%%%
%%%%%%%%%%%%%%%%%%%%%%%%%%%%%%%%%%%%%%%%%%%%%%%%%%%%%%%%%%%%%%%%%%%%%%%%%%%%%%%%%%%%%%%%%%%%%%%%%%
\end{document}